\documentclass{aastex62}
\usepackage{enumerate}

\shorttitle{}
\shortauthors{Li et al.}

\usepackage{graphicx}

\begin{document}

\title{A $Herschel$ mapping of [C {\sc ii}], [O {\sc i}] and [O {\sc iii}] lines from the circumnuclear region of M31}
\author{Zongnan Li}
\email{lzn@smail.nju.edu.cn}
\author{Zhiyuan Li}
\email{lizy@nju.edu.cn}
\affiliation{School of Astronomy and Space Science, Nanjing University, Nanjing 210023, China}
\affiliation{Key Laboratory of Modern Astronomy and Astrophysics, Nanjing University, Nanjing 210023, China}
\author{Matthew W. L. Smith}
\affiliation{School of Physics \& Astronomy, Cardiff University, The Parade, Cardiff CF24 3AA, UK}
\author{Yu Gao}
\affiliation{Department of Astronomy, Xiamen University, Xiamen, Fujian 361005, China}
\affiliation{Purple Mountain Observatory and Key Laboratory for Radio Astronomy, Chinese Academy of Sciences, Nanjing 210023, China}

\begin{abstract}

The circumnuclear region of M31, consisting of multiphase interstellar medium, provides a close-up view of the interaction of the central supermassive black hole and surrounding materials. Far-infrared (FIR) fine structure lines and their flux ratios can be used as diagnostics of physical properties of the neutral gas in this region. 
Here we present the first FIR spectroscopic mapping of the circumnuclear region of M31 in [C {\sc ii}] 158$\,\micron$, [O {\sc i}] 63$\,\micron$ and [O {\sc iii}] 88$\,\micron$ lines with the $Herschel~Space~Observatory$, covering a $\sim500\times500$ pc (2$'\times2'$) field. Significant emissions of all three lines are detected along the so-called nuclear spiral across the central kpc of M31. 
The velocity field under a spatial resolution of $\sim$ 50 pc of the three lines are in broad consistency and also consistent with previous CO(3--2) line observations in the central region. Combined with existing [C {\sc ii}] and CO(3--2) observations of five other fields targeting on the disk, we derived the radial distribution of [C {\sc ii}]/CO(3--2) flux ratio, and found that this ratio is higher in the center than the disk, indicating a low gas density and strong radiation field in the central region. We also found that the [C {\sc ii}]/FIR ratio in the central region is 5.4 ($\pm$ 0.8) $\times$ 10$^{-3}$, which exhibits an increasing trend with the galactocentric radius, suggesting an increasing contribution from old stellar population to dust heating towards the center.

\end{abstract}


\section{Introduction}

Galactic circumnuclear regions, in which the multiphase interstellar medium (ISM) and various stellar populations are actively interacting with the supermassive black holes (SMBHs), are crucial to our understanding of the accretion and feedback of SMBHs, and in turn the global evolution of the host galaxies. 
As the closest external spiral galaxy, the Andromeda galaxy (M31) gives us a unique opportunity to study the circumnuclear environment in a close-up view, thanks to its proximity \citep[D $\approx$ 785 kpc, where 1$\arcsec$ = 3.8 pc,][]{McConnachie 2005}, appropriate inclination ($\sim 77^\circ$) and low line-of-sight extinction \citep[$A\rm_V \lesssim$ 1,][]{Dong 2016}. 
The most conspicuous structure of this environment is the so-called nuclear spiral, which consists of spiral-like filaments across the central few hundred parsecs of M31. Both optical emission \citep{Jacoby 1985} and extinction \citep{Dong 2016}, and infrared (IR) emission \citep{Li 2009} from the nuclear spiral indicate that it is composed of neutral, dusty gas clouds with ionized outer layers. 
The investigation of dust \citep{Draine 2014} and molecular gas \citep[e.g.][]{Melchior 2013, Li 2019, Li 2020} has revealed that the kinetic temperature is higher in the nucleus than in the disk. 
However, the ionization/excitation mechanism of the nuclear spiral remains a long-standing puzzle, given the weak nuclear activity \citep{Li 2011} and the absence of on-going star formation \citep{Brown 1998, Groves 2012, Dong 2015} in this region. 
\\

Far infrared (FIR) atomic/ionic fine-structure lines, such as [C {\sc ii}] 158$\,\micron$, [O {\sc i}] 63$\,\micron$ and [O {\sc iii}] 88$\,\micron$, arise from multiple ISM phases, and thus can be used as diagnostics of the physical conditions of the circumnuclear environment. 
Specifically, due to the moderate ionization potential of carbon (11.26 eV), 
[C {\sc ii}] 158 $\micron$ line is present in low-density diffuse ionized gas, neutral photodissociation regions (PDRs) and also diffuse molecular clouds \citep{Petuchowski 1993, Heiles 1994, Abel 2006}. On the other hand, [O {\sc i}] 63 $\micron$ line is only found in neutral gas (PDRs and molecular clouds) since its ionization potential is 13.62 eV, very close to that of hydrogen; while [O {\sc iii}] 88 $\micron$ is exclusively found in ionized gas because the production of O$^{++}$ requires energetic photons (35.12 eV). 
These lines are also widely used as star formation rate (SFR) tracers \citep[e.g.][]{De Looze 2014} since they are from the ISM that fuels and closely interacts with star formation. Among them, [C {\sc ii}] 158$\,\micron$ is most often investigated since it is typically the most luminous line in galaxies \citep[e.g.,][]{Stacey 1991}. The relationship of SFR surface density and [C {\sc ii}] intensity has been extensively investigated in nearby galaxies \citep{Herrera-Camus 2015}, such as M51 \citep{Pineda 2018}, and also in M31 disk \citep{Kapala 2015}. These studies all found tight correlations between the two quantities. However, since the circumnuclear region of M31 is lack of current star formation \citep[e.g.][]{Dong 2015}, we will not further discuss this correlation in this paper.
\\

Moreover, [C {\sc ii}] 158$\,\micron$ usually acts as the main coolant for low density neutral gas (the bulk of neutral gas) due to its lower excitation energies and critical densities ($\Delta E/k$ $\sim$91 K and $n\rm_{crit} \sim$ 3$\times10^3$ cm$^{-3}$), while [O {\sc i}] 63 $\micron$ is dominant in high density and temperature regime ($\Delta E/k$ $\sim$228 K; $n\rm_{crit}\sim$8.5$\times$10$^5$ cm$^{-3}$). The neutral ISM is mainly heated by photoelectrons ejected by dust grains absorbing far-ultraviolet (FUV) photons within the range 6-13.6 eV \citep{Draine 1978}. The energy of the heated ISM is later taken away by collisional excited FIR emission lines, and the gas is thus cooled. On the other hand, the dust grains can be heated by photons of nearly all energies (mostly 0.01-13.6 eV) and then re-radiated in total infrared \citep[TIR, 8-1000 $\micron$;][]{Sanders 1996} band. Thus, the [C {\sc ii}]/TIR ratio can be an approximation of the theoretical photoelectric heating efficiency, $\epsilon\rm_{PE}$, defined as the photoelectric heating rate divided by the dust grain photon absorption rate \citep{Tielens 1985}. 
Investigations of these lines could give us insights about the physical properties of the ISM, in particular the M31 center.
\\

The first [C {\sc ii}] 158$\micron$ observations of M31 \citep{Mochizuki 2000} was made with the {\it Infrared Space Observatory} \citep[$ISO$;][]{Kessler 1996}, resulting in a strip map of 31 positions with a spatial resolution of $\sim$70$\arcsec$. \cite{Mochizuki 2000} detected [C {\sc ii}] flux in the central region at the level of 1$\times 10^{-5}$ erg s$^{-1}$ cm$^{-2}$ sr$^{-1}$ and found a [C {\sc ii}]/100 $\micron$ flux ratio of  6$\times10^{-3}$, higher than that found in the Milky Way, which the author interpreted as due to the higher gas column density of the Galactic center. 
More recent observations \citep{Kapala 2015} carried out with {\it Herschel Space Observatory} \citep{Pilbratt 2010} targeting five regions in M31 disk have mapped the [C {\sc ii}] lines with much higher resolution (12$\arcsec$) and found the [C {\sc ii}]/TIR ratio increases with radius. 
Using the same set of data, \cite{Kapala 2017} later claimed that since TIR emission involves the contribution from all photons that can heat dust, not just the FUV photons capable of ejecting photoelectrons, the [C {\sc ii}]/FUV$\rm_{att}$ ratio could be a better proxy of $\epsilon\rm_{PE}$, where FUV$\rm_{att}$ stands for the attenuated FUV emission. 
\\

The circumnuclear region of M31 is a relatively small reservoir of neutral gas. To date, there is no unambiguous detection of circumnuclear atomic hydrogen reported. An upper limit of $10^6$ M$_\odot$ was placed on the HI mass in the central 500 pc \citep{Brinks 1984}, and a more recent survey by \cite{Braun 2009} yielded no substantial detection in the central kpc. Only until recently, molecular gas has been mapped in this region with sensitive IRAM 30m CO(2--1) \citep{Melchior 2013} and JCMT CO(3--2) observations \citep{Li 2019}, which suggest a small amount of molecular gas ($\sim3.7 \times 10^5$ M$_\odot$).
In this work we present $Herschel$ spectroscopic observations of the [C {\sc ii}] 158 $\micron$, [O {\sc i}] 63 $\micron$ and [O {\sc iii}] 88 $\micron$ lines in the central 2 arcmin of M31, which is the first mapping of atomic phase gas in this region with a resolution comparable to molecular gas and dust observations. We make maps and analyze the flux ratios of these lines to study the ISM properties of the nuclear spiral and compare the results with that of the disk. 
The paper is organized as follows. The observations and data preparation are described in Section 2. The results are presented in Section 3, including the morphology and kinematics of the lines, and the analysis of their ratios. The discussion and implication of the results are presented in Section 4. A brief summary is given in Section 5.
\\

\section{Observations and data preparation}

The observations of [C {\sc ii}] 158$\,\micron$, [O {\sc i}] 63$\,\micron$ and [O {\sc iii}] 88$\,\micron$ lines were carried out during 2012-6-19 and 2012-6-20 for a total time of 9.9 hours (obsid: 1342247148, 1342247149; PI: Z. Li) with $Herschel~Space~Observatory$. The data were obtained with Photodetector Array Camera and Spectrometer \citep[PACS;][]{Poglitsch 2010} using the 3 $\times$ 3 raster mode covering a 120$\arcsec \times 120\arcsec$ ($\sim500\times500$ pc) area of M31 nuclear spiral. This field is centering on ([RA,DEC]=[10.691149, 41.275203]), $\sim$100 pc northeast of the M31 center ([RA,DEC]=[10.684793, 41.269065]). We adopted the unchopped grating scan mode, with an off-source field chosen at an inter-arm region $\sim500\arcsec$ northwest of the M31 nucleus. The wavelength, angular and velocity resolution of these lines are given in Table \ref{tab1}. 
\\

The data were reduced using the Herschel Interactive Processing Environment \citep[HIPE,][]{Ott 2010} version 14.0 standard pipeline (SPG 14). We made use of the level 2 data from the Herschel Science Archive (HSA), which are fully reduced and flux calibrated. We then applied the script provided by $Herschel$ named ``FittingPacsCubes\_mappingI.py", which is suitable for PACS spectroscopy data cube analysis, to process the data and fit the spectra. The main steps are summarized below.
\\

We started with the interpolated cube with a pixel size of 3$\arcsec$ since it not only satisfies the Nyquist sampling requirement for all three lines but also is smoother than the recommended projected cube. 
First, we cropped the noisy end of the spectra to avoid contamination from the spectral edges. Next, we used a first-order polynomial (second-order for the [O {\sc i}] spectrum, since the signal is relatively weak and the baseline is not as stable as the other two lines) plus a Gaussian to fit each spectrum, and obtained the peak, center and width of the line. We then eliminated the spaxels with peak signal-to-noise ratio (SNR) smaller than 3 and obtained the integrated intensity and velocity maps of all three lines using the masked cube. For comparison of CO(3--2) and [C {\sc ii}] emission, we smoothed them to a common spatial resolution and regridded them to a pixel size of 15$\arcsec$. 
The flux is converted to erg s$^{-1}$ cm$^{-2}$ sr$^{-1}$, and the velocity reference frame is Local Standard of Rest (LSR). Throughout this paper, we adopt a systemic velocity of M31 of -300 km~s$^{-1}$ \citep{van den Bergh 1969}, a position angle of 38$^\circ$ and an inclination angle of 77$^\circ$ for the M31 disk \citep{McConnachie 2005}, and an inclination angle of 45$^\circ$ for the central 1 kpc \citep{Ciardullo 1988}.
\\

\begin{figure}
\centering
\includegraphics[width=7in]{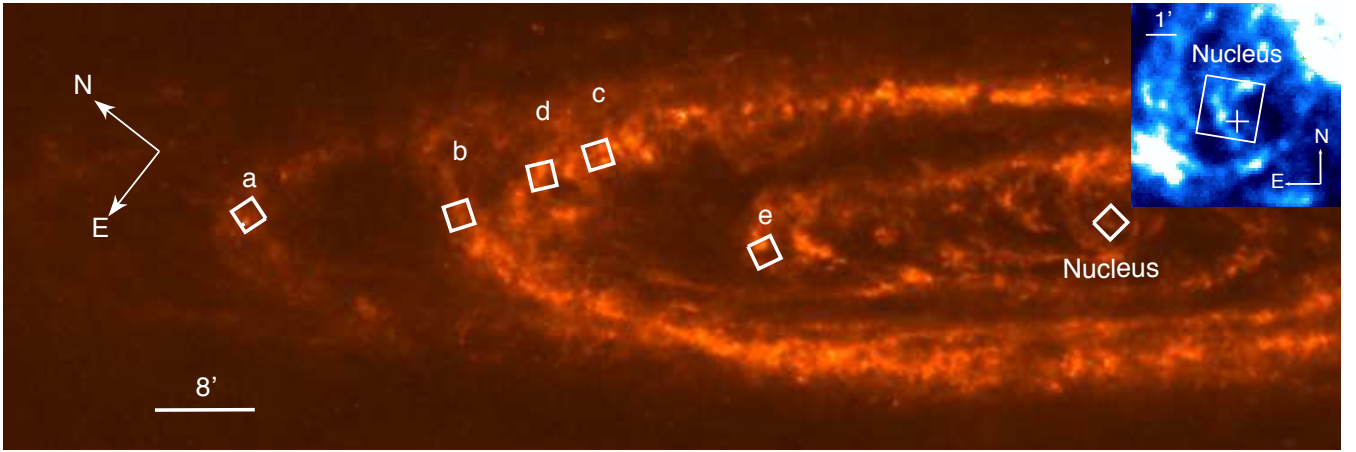}
\caption{ The footprint of the five SLIM fields (a-e) and the Nucleus field (`Nucleus') overlaid on the $Herschel$/SPIRE 250 $\micron$ image of M31. The insert is a zoom-in of the nuclear region, where the nuclear spiral is located. The white cross in the insert marks the center of M31. 
\label{fig:fig1}}
\end{figure}

We retrieved several ancillary data sets to assist our analysis: (1) The CO(3--2) map of the central kpc of M31 obtained from the James Clerk Maxwell Telescope (JCMT) with HARP raster mode \citep{Li 2019}, taken with a resolution of 15$\arcsec$ and a typical RMS of 3.5 mK in a 13 km s$^{-1}$ channel. (2) $Herschel$ PACS imaging in [C {\sc ii}] 158 $\micron$ 
and optical integral-field spectroscopy of five fields in M31 disk from the Survey of Lines In M31 \citep[SLIM,][]{Kapala 2015}. These fields are reduced in the same way as described above. (3) The CO(3--2) observations of the five SLIM fields from a JCMT large program HARP and SCUBA-2 High-Resolution Terahertz Andromeda Galaxy Survey \citep[HASHTAG;][M. Smith et al. in preparation]{Li 2020}. (4) Galaxy Evolution Explorer (GALEX) FUV (1350-1750 $\rm\r{A}$) map of M31 \citep{Thilker 2005}. (5) The dust mass surface density map of the whole galaxy \citep{Draine 2014}, obtained with spectral energy distribution (SED) fitting of infrared emission. (6) Total far infrared (FIR) luminosity map from \cite{Ford 2013}, which was produced by integrating over six frequency bands, including $Spitzer$ MIPS 70 $\micron$, $Herschel$ PACS and SPIRE 100, 160, 250, 350, 500 $\micron$. The footprint of the five SLIM fields as well as the Nucleus field are shown in Figure \ref{fig:fig1}.
\\

\begin{figure}
\centering
\includegraphics[width=3in]{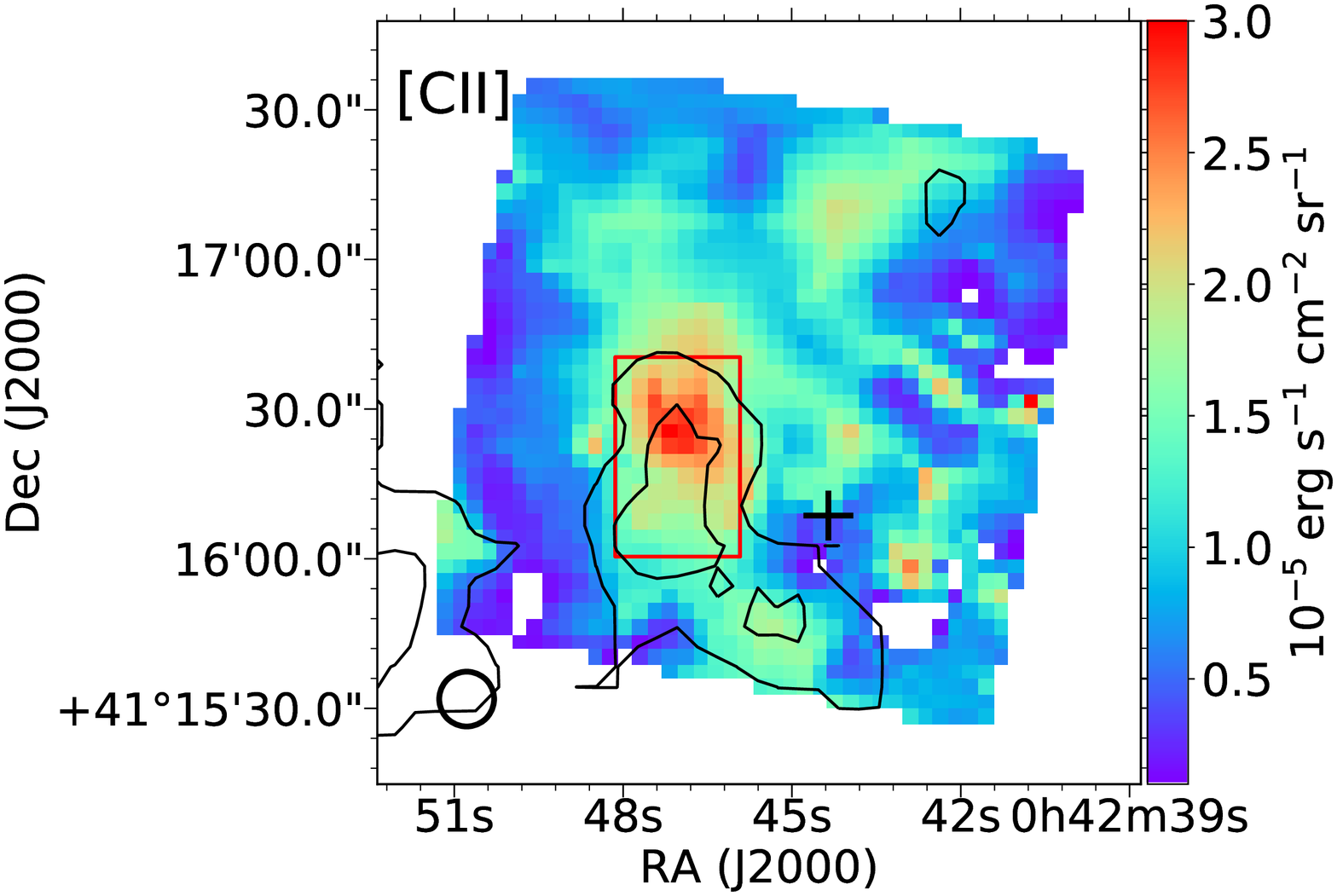}
\includegraphics[width=3in]{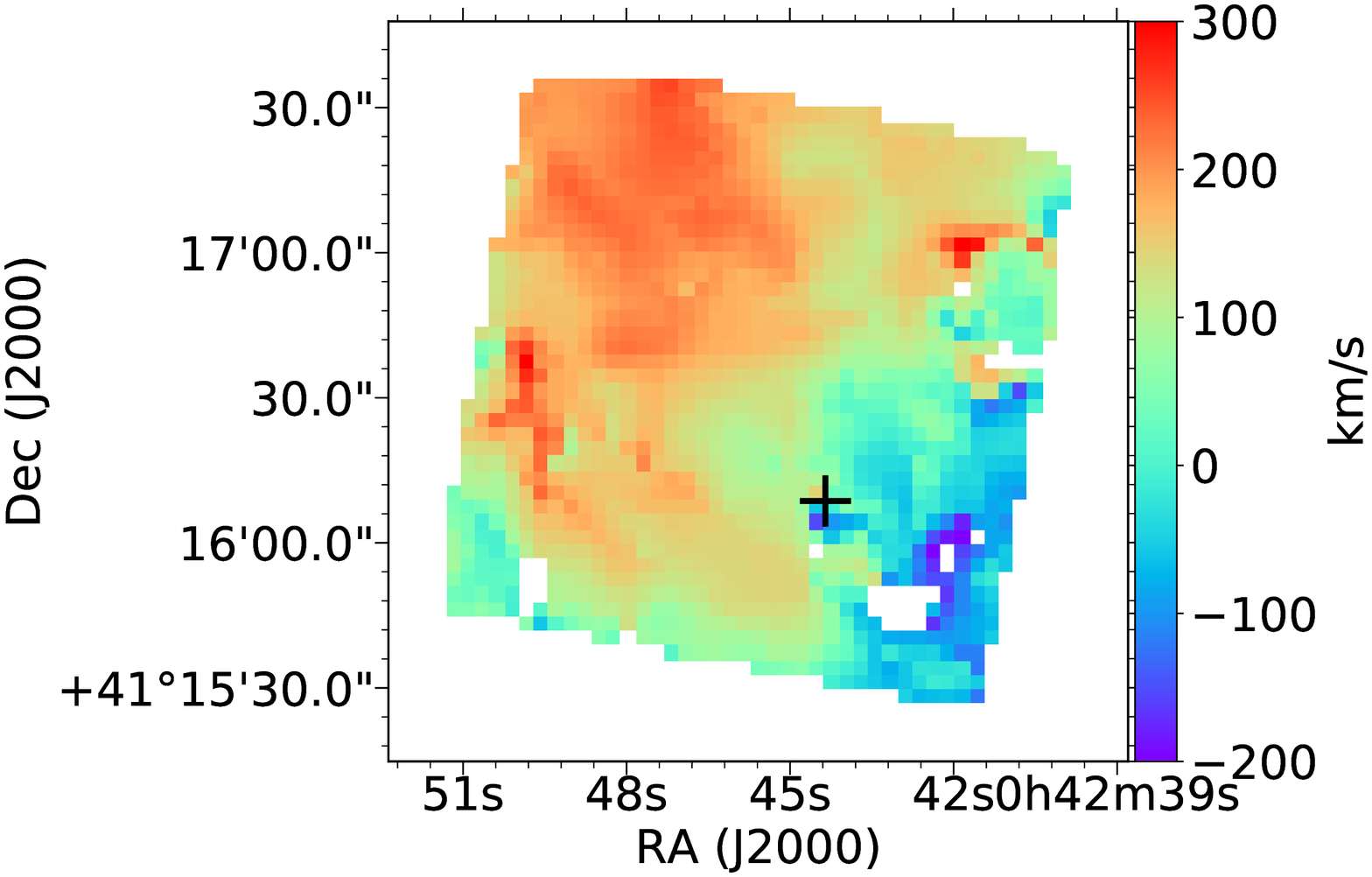}
\includegraphics[width=3in]{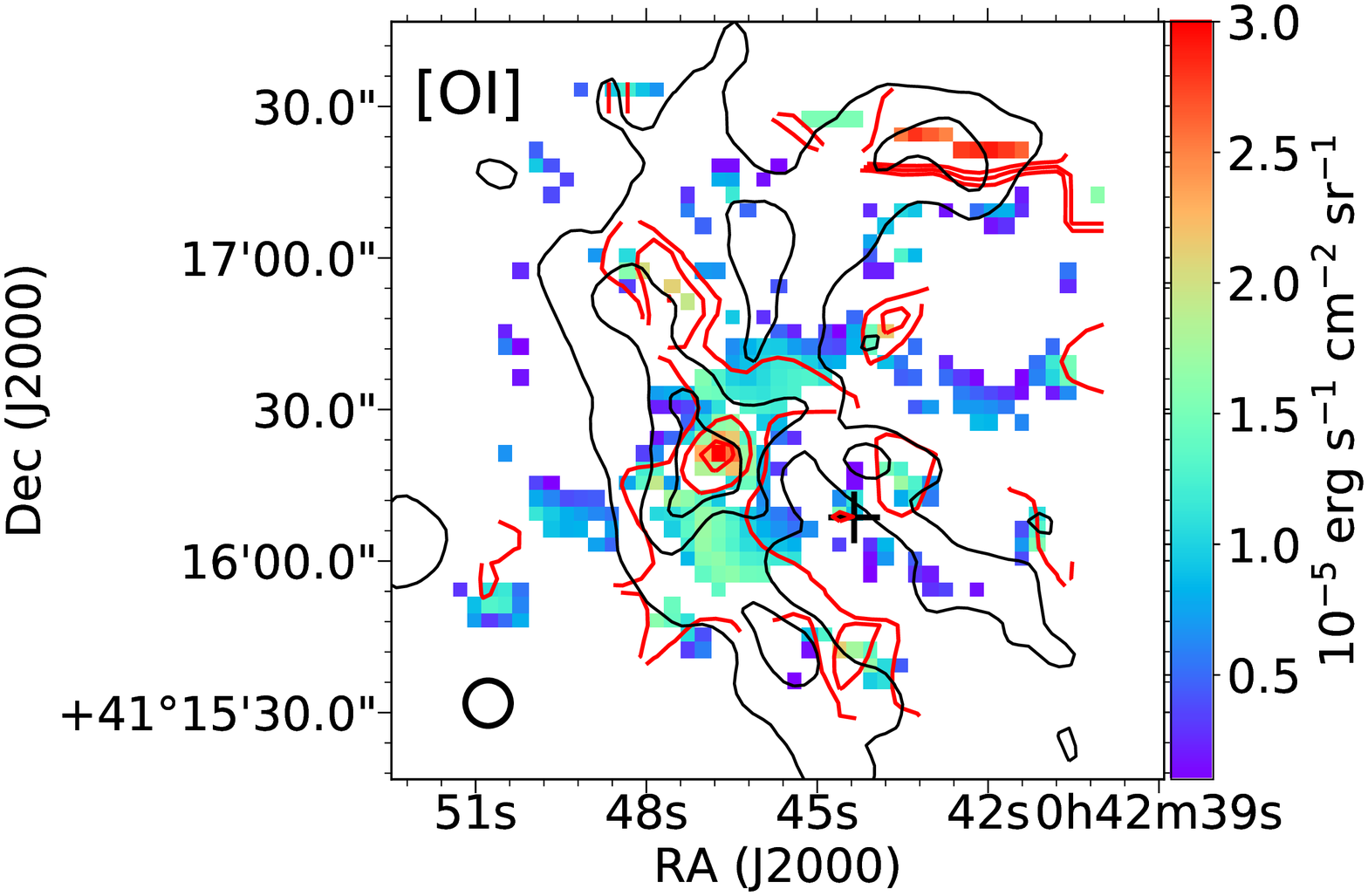}
\includegraphics[width=3in]{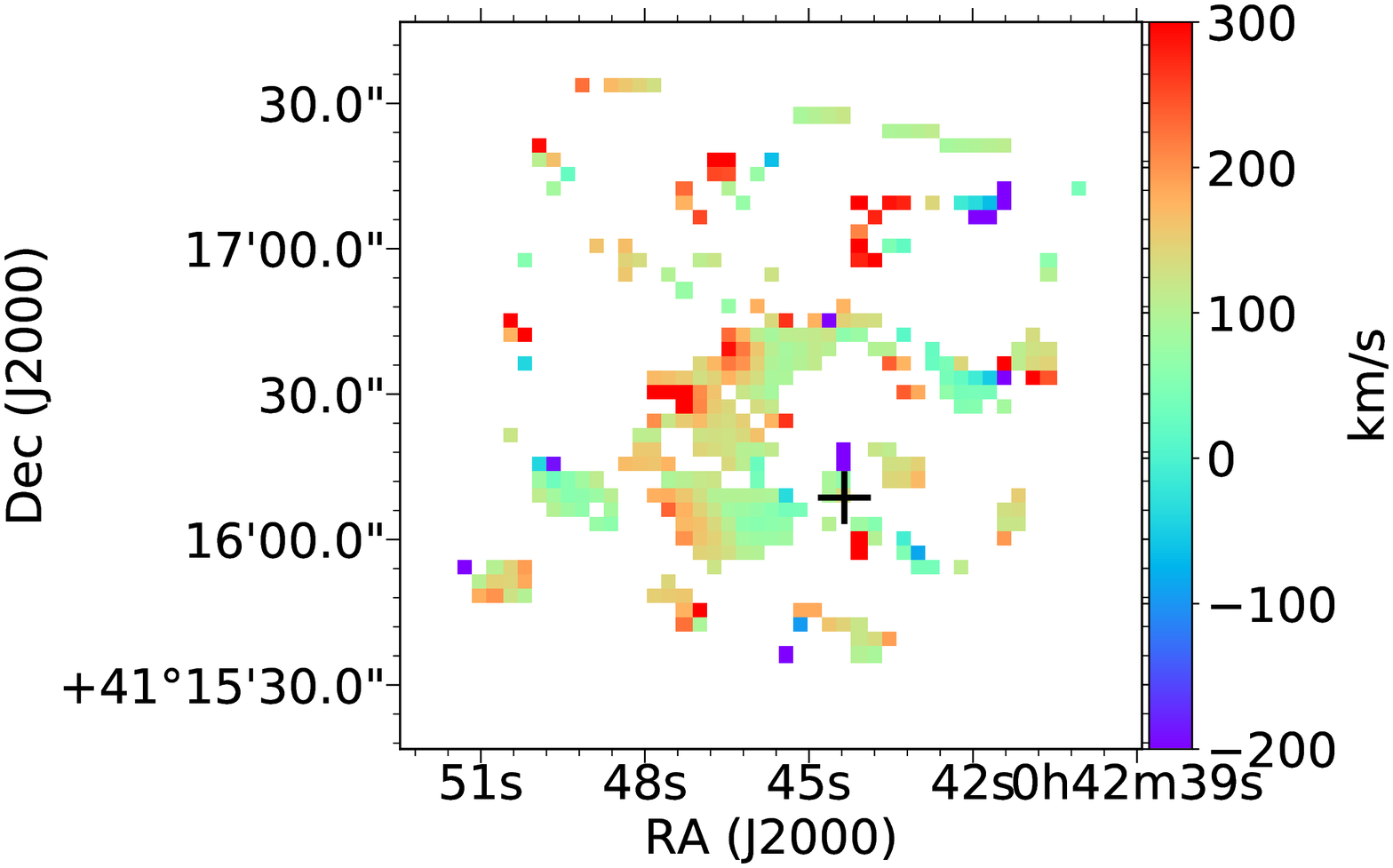}
\includegraphics[width=3in]{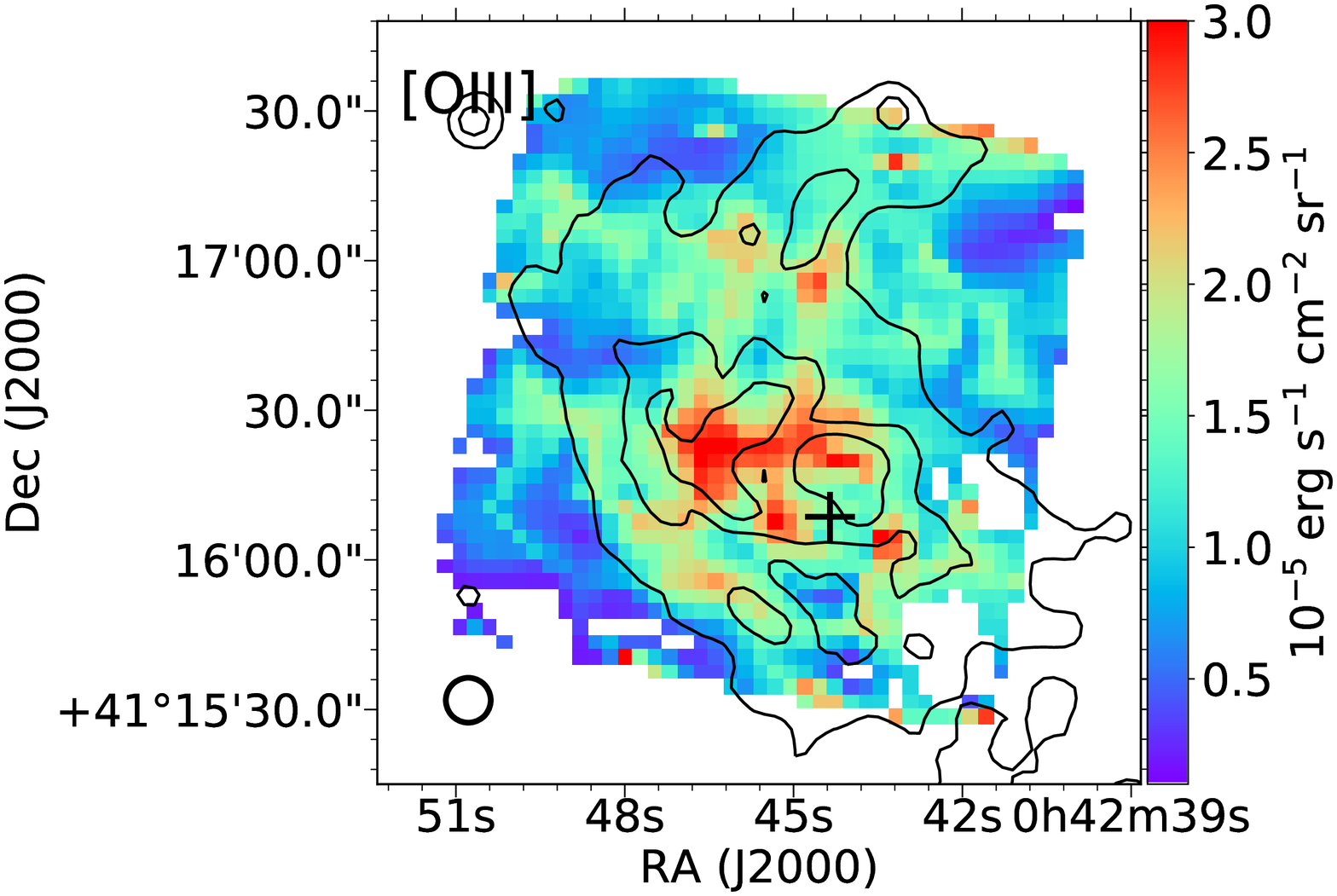}
\includegraphics[width=3in]{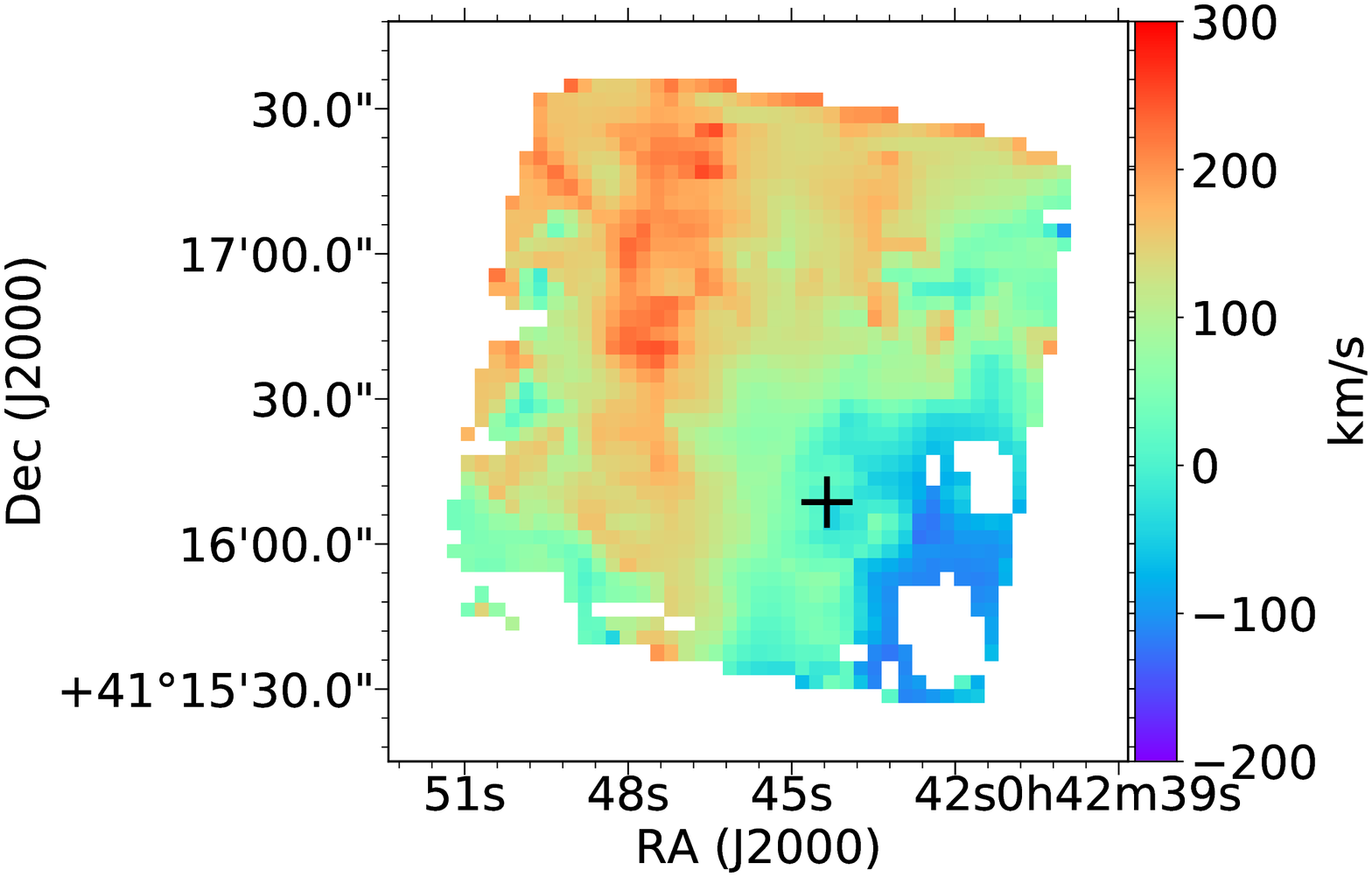}
\caption{Integrated intensity map (left) and velocity map (right) of [C {\sc ii}], [O {\sc i}] and [O {\sc iii}]. [C {\sc ii}] is overlaid with CO(3-2) contours from \cite{Li 2019}, with levels 0.07 and 0.35 K km/s. [O {\sc i}] is overlaid on $Herschel$/PACS 70$\micron$ emission from \citep{Groves 2012} (black). The red curves are smoothed [O {\sc i}] contours to guide the eye. [O {\sc iii}] is overlaid on H$\alpha$ map \citep{Devereux 1994}. The black cross marks the M31 center. The red rectangle in the first panel outlines the region where spectra are being averaged, as shown in Figure \ref{fig:fig3}. For clarity, we did not show it in the [O {\sc i}] and [O {\sc iii}] maps. This particular region is chosen to analyze the spectra to ensure a sufficient SNR in all four lines. The systemic velocity of M31 has been subtracted in the velocity map. The black circle in the lower left corner indicates the resolution of each map.
\label{fig:fig2}}
\end{figure}


\begin{deluxetable*}{ccccc}
\tablecaption{Lines observed with PACS spectrometer.}
\tablenum{1}
\tablewidth{2pt}
\tablehead{
\colhead{Species} & \colhead{Rest wavelength} & \colhead{Velocity resolution\tablenotemark{(a)}} & \colhead{Angular resolution} & \colhead{Flux\tablenotemark{(b)}}  \\
& \colhead{($\micron$)} & \colhead{(km s$^{-1}$)} & \colhead{($\arcsec$)} & \colhead{(10$^{-5}$ erg s$^{-1}$ cm$^{-1}$ sr$^{-1}$)}
}
\startdata
[O {\sc i}] & 63.18 & 98 &9& 1.06 $\pm$ 0.26 \\
\[[O {\sc iii}] & 88.36 & 121 &9& 1.85 $\pm$ 0.17 \\
\[[C {\sc ii}] &157.74 & 239 &11& 2.03 $\pm$ 0.08 \\
\enddata
\tablecomments{(a) The velocity and angular resolution are from PACS observers' manual 2010 (http://herschel.esac.esa.int/Docs/PACS/pdf/pacs\_om.pdf). (b) The average intensity of the lines measured within the red rectangle in Figure \ref{fig:fig2}. The error represents the typical intensity error of each spaxel. \label{tab1}}
\end{deluxetable*}

\section{Results}

\subsection{Morphology and kinematics}

The integrated intensity and velocity maps of the three lines are shown in Figure \ref{fig:fig2}. As shown in the upper left panel, the [C {\sc ii}] integrated intensity (color map) is in broad consistency with the CO(3-2) emission (contours) in the southern half of the nuclear spiral, while CO(3--2) is almost absent in the northern half, which is mainly due to the weak CO(3--2) emission in this region \citep{Li 2019}. 
[O {\sc i}] line is weak compared to the other two lines. Though the emission seems patchy due to the low SNR, there are still detections in the nuclear spiral region, and the morphology is roughly consistent with the dust emission. 
The [O {\sc iii}] emission has fibrous structure and broadly matches with the H$\alpha$ emission \citep{Devereux 1994}, which is obtained with the Case Western Burrell Schmidt telescope at Kitt Peak National Observatory. 
The line-of-sight velocity fields of the three lines are shown in the right panels of Figure \ref{fig:fig2}. The velocity patterns are similar, all indicating that the nuclear spiral structure is redshifted for $\sim$200 km s$^{-1}$ with respect to M31's systemic velocity, consistent with the CO(3--2) velocity obtained with JCMT \citep{Li 2019}.
\\

\begin{figure}
\centering
\includegraphics[width=3.5in]{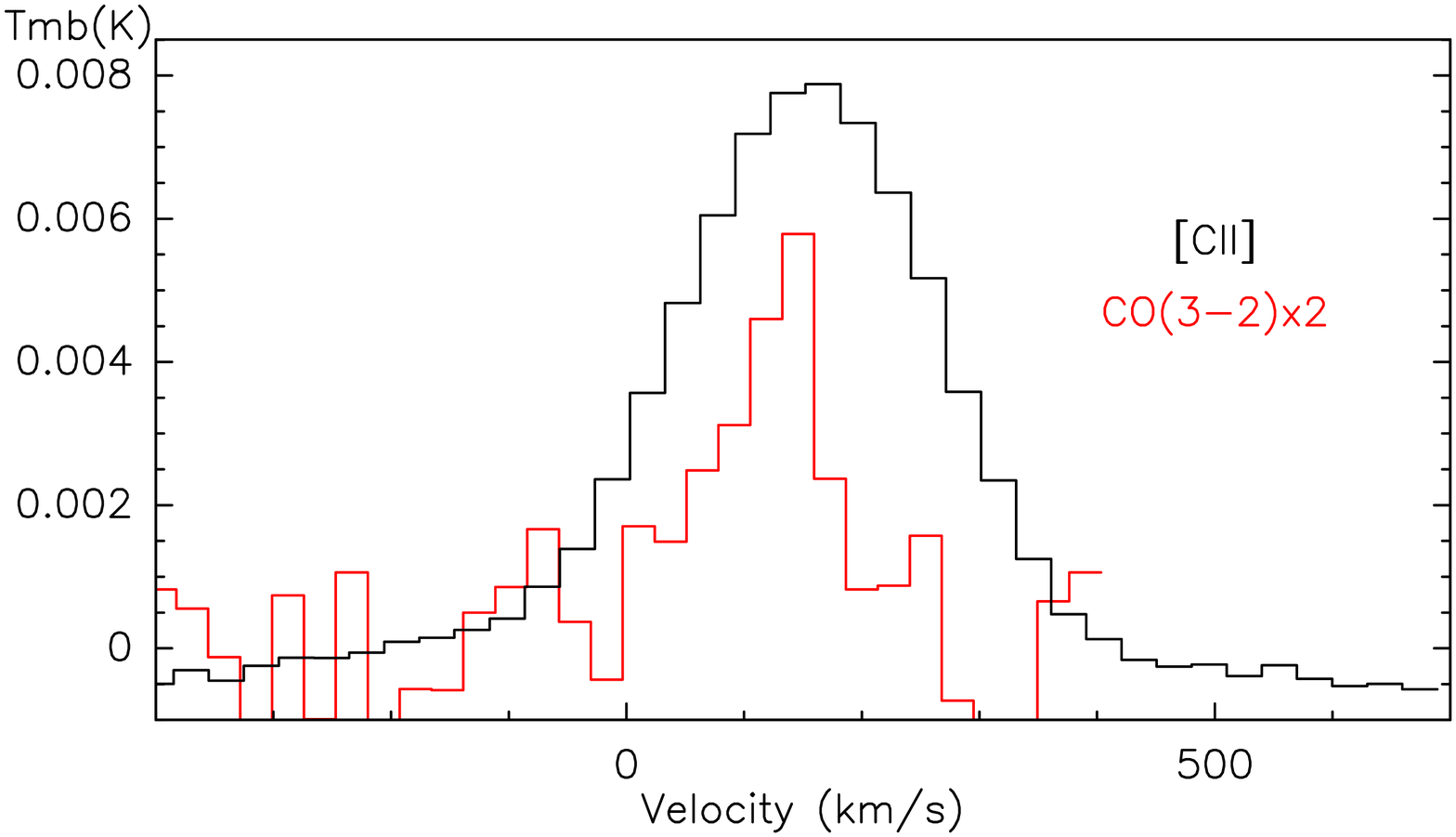}
\includegraphics[width=3.5in]{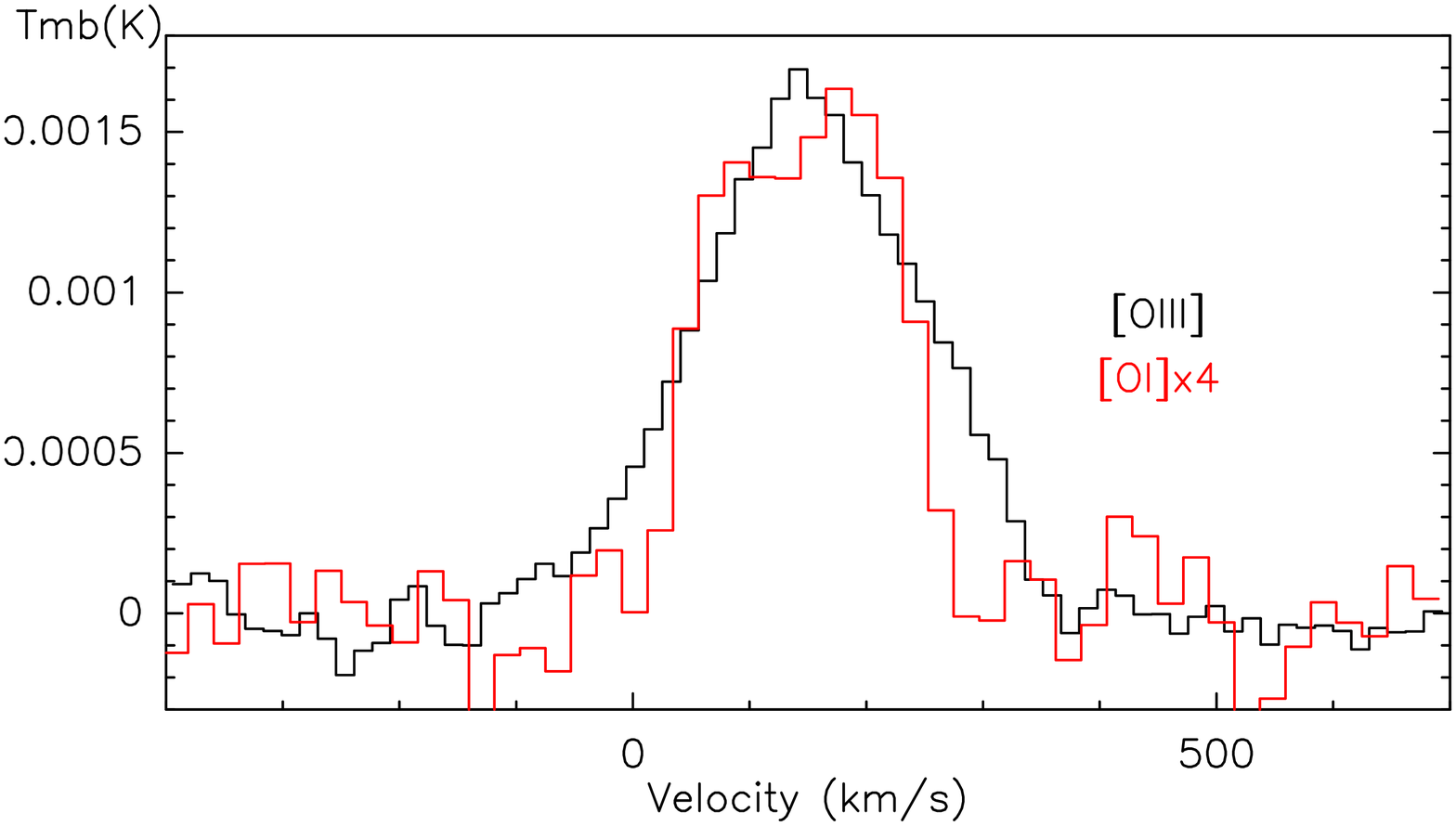}
\caption{Left: Averaged spectra of CO(3--2) and [C {\sc ii}]. Right: Averaged [O {\sc i}] and [O {\sc iii}] spectra. These spectra are averaged over the rectangle region as shown in the first panel of Figure \ref{fig:fig2} to ensure a sufficient SNR. The systemic velocity of M31 has been subtracted.
\label{fig:fig3}}
\end{figure}

Averaged spectra of the the CO(3--2) and [C {\sc ii}] lines are shown in Figure \ref{fig:fig3} left panel, and that of [O {\sc i}] and [O {\sc iii}] lines are shown in the right panel. To ensure a sufficient SNR, we only included the spaxels in the red rectangle shown in Figure \ref{fig:fig2}, where the emission is strong for all the four lines. 
The average flux of each line measured within the rectangle is listed in Table \ref{tab1}. For the weak [O {\sc i}] line, we further derived a 3 $\sigma$ upper limit, $\rm \sim 2\times 10^{-6}~erg~cm^{-2}~s^{-1}~sr^{-1}$, over spaxels with SNR $<$ 3 throughout the map. 
After Gaussian fitting and deconvolution, we found the CO(3--2) and [C {\sc ii}] line are consistent in central velocity ($\sim$150 km s$^{-1}$) and velocity dispersion ($\sim$ 120 km s$^{-1}$). 
The central velocity of the [O {\sc i}] and [O {\sc iii}] lines are also consistent with each other, while the [O {\sc iii}] line is broader than [O {\sc i}] by $\sim$ 30 km s$^{-1}$. The relatively large difference in the velocity dispersion is mainly caused by the excess in the [O {\sc iii}] line at around 270 km s$^{-1}$, which is unseen in [O {\sc i}]. Since we averaged spectra from the same region (the red rectangle in Figure \ref{fig:fig2}), this excess could be an ionized gas component along the line-of-sight that is not traced by [O {\sc i}]. 
Under a relatively high velocity resolution, the [O {\sc i}] line exhibits two velocity components, centering at $\sim$190 km s$^{-1}$ and $\sim$80 km s$^{-1}$, respectively. This is best understood as the [O {\sc i}] line arising from two different arms of the nuclear spiral, as shown in the extinction map obtained with multi-band HST observations \citep[][Figure 3 therein]{Dong 2016}. 
\\

\subsection{Flux ratios}


[C {\sc ii}] is the dominant coolant of neutral gas, which is mainly heated by UV photons with energy range 6-13.6 eV (equivalent to 912-2067$\rm\r{A}$). On the other hand, CO is mainly formed in dense H$_2$ gas and shielded by dust grains, which can be photodissociated by absorbing photons near 1000 $\rm\r{A}$. Clouds with higher metallicity and gas density will be able to shield the CO molecule more efficiently and thus enhance the CO emission. Meanwhile, FUV photons can both increase [C {\sc ii}] and decrease CO emission. 
As a result, the [C {\sc ii}]/CO(3--2) flux ratio varies with metallicity, gas density, and FUV flux \citep[e.g.][]{Rollig 2006}. 
\\

\begin{figure}
\centering
\includegraphics[width=3.5in]{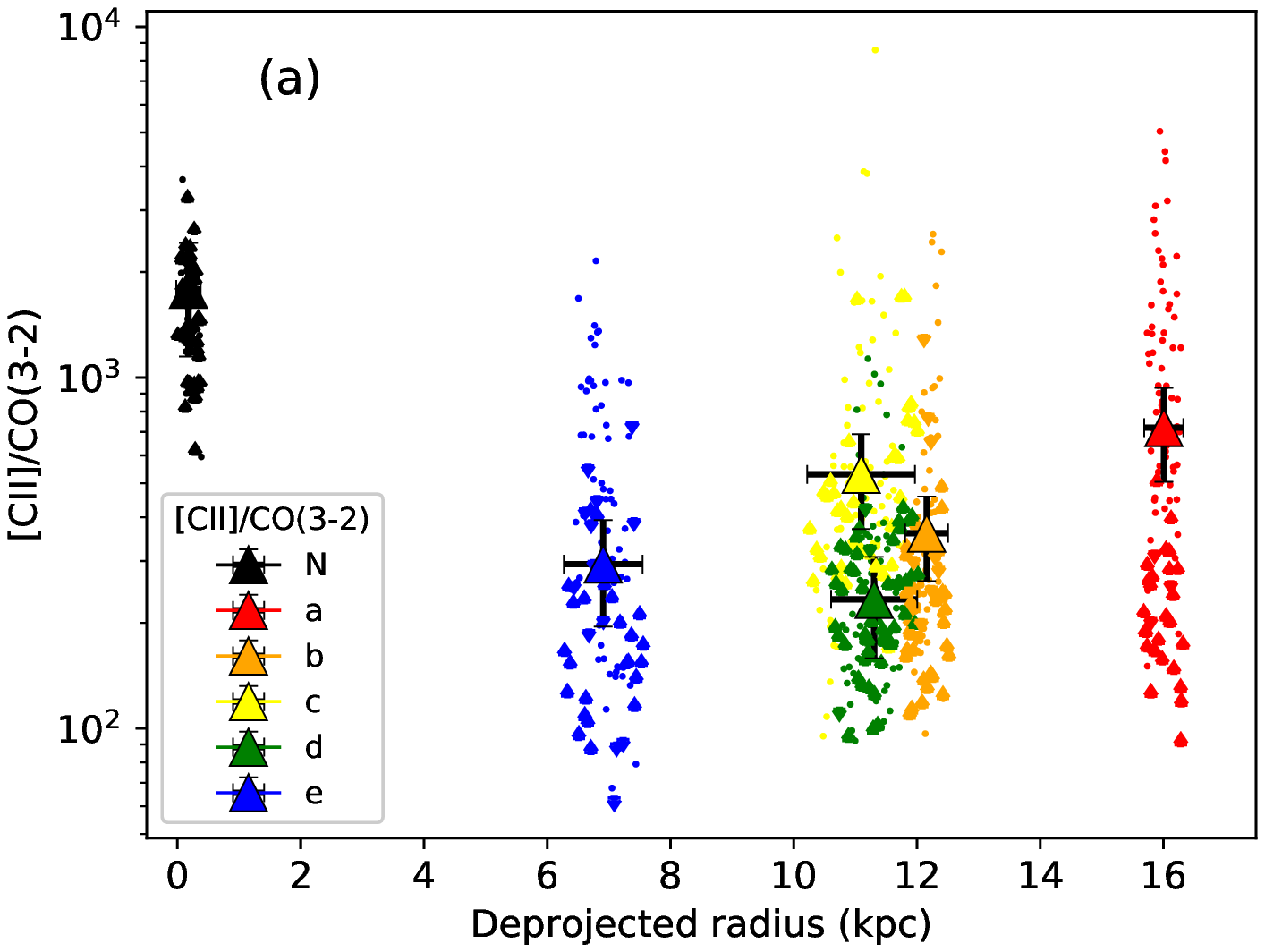}
\includegraphics[width=3.5in]{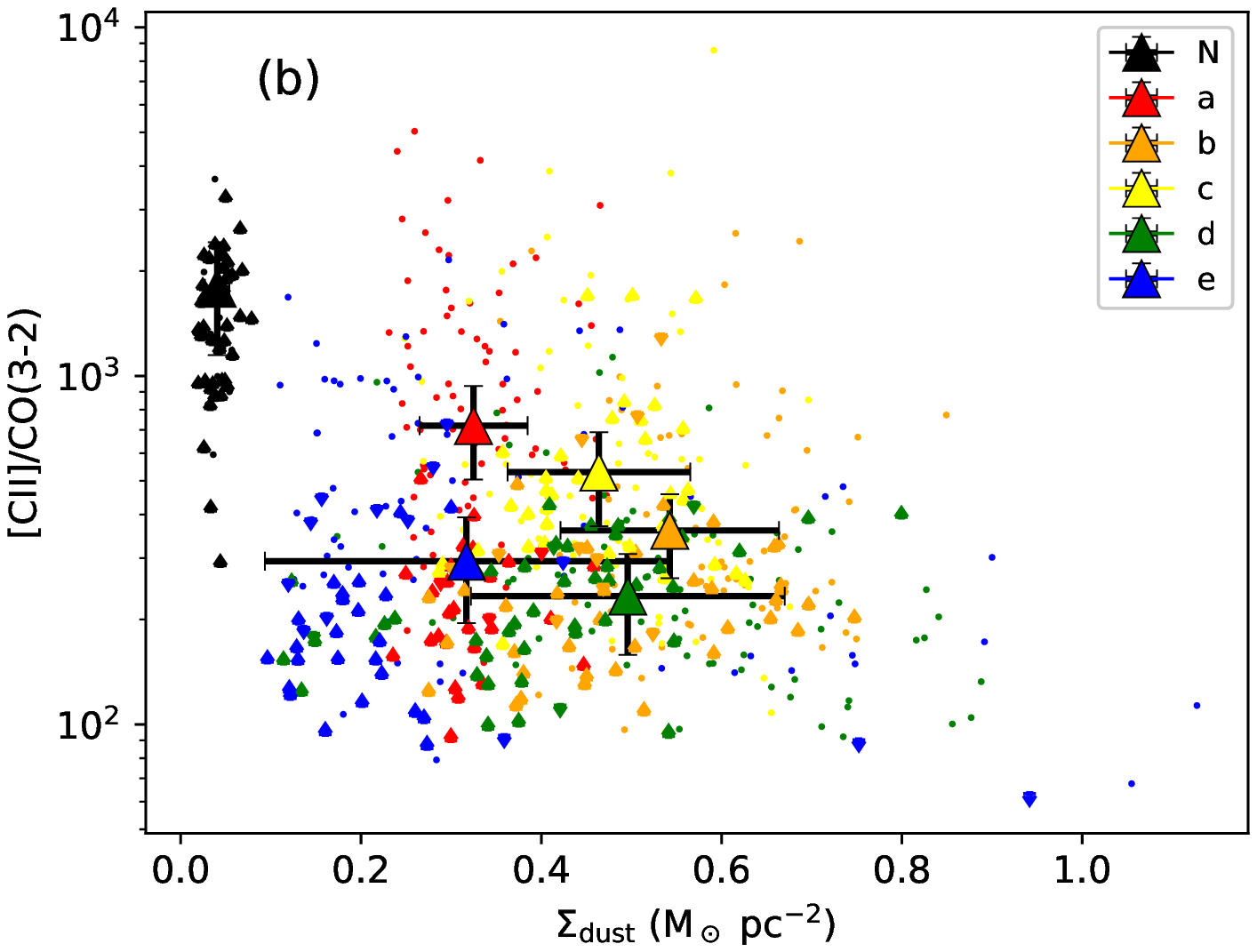}
\includegraphics[width=3.5in]{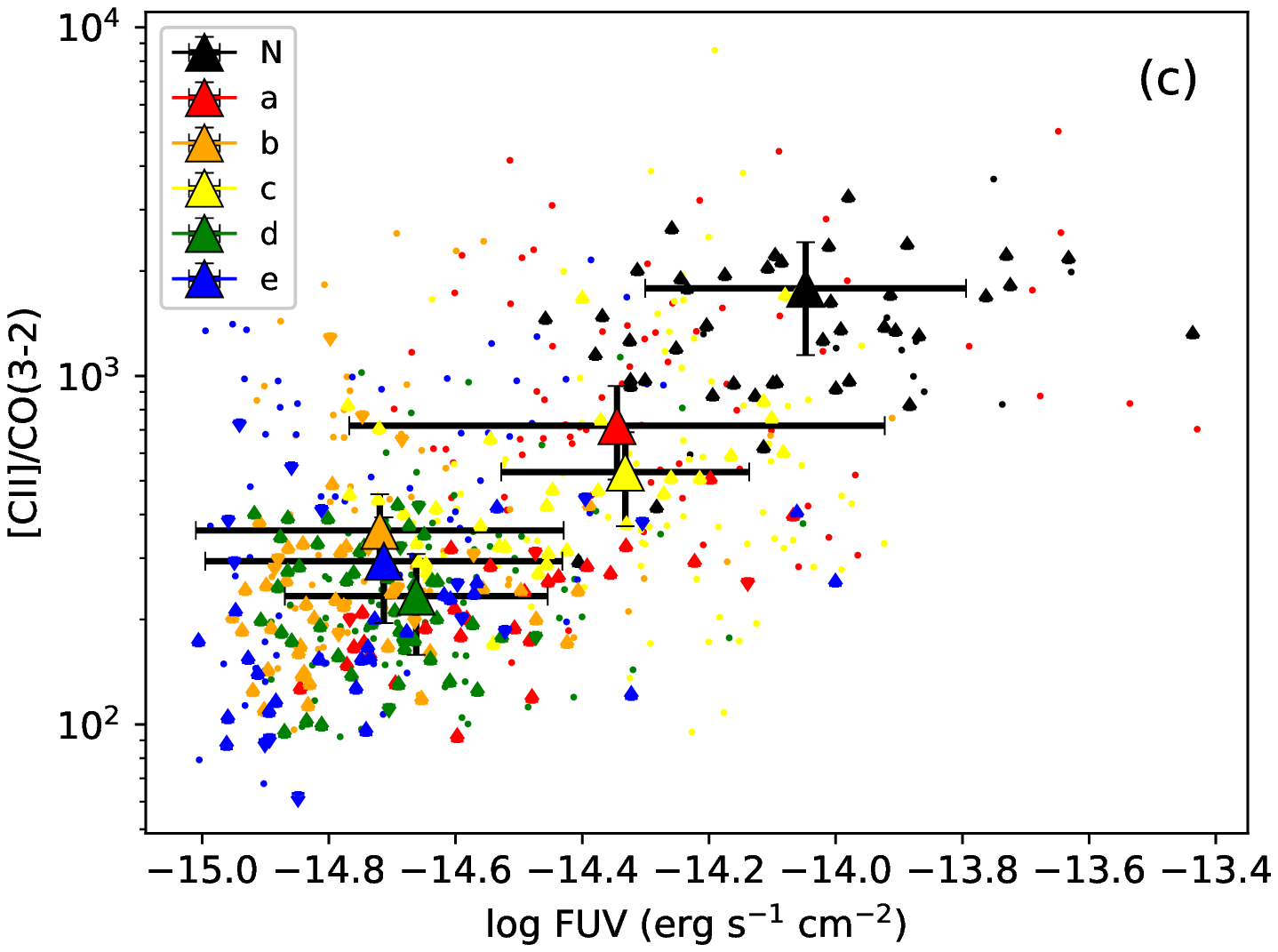}
\caption{(a) The radial distribution of [C {\sc ii}]/CO(3--2) intensity ratio. The dots are data points with both CO(3--2) and [C {\sc ii}] detections, upward and downward triangles represent upper and lower limits of this ratio. The mean ratio (triangles) are calculated using software ASURV. The five SLIM fields (denoted by a-e) and Nucleus field (`N') are separated by different colors. The vertical and horizontal error bars represent the uncertainty of the mean ratio and the radial coverage of each field. (b) [C {\sc ii}]/CO(3--2) ratio vs. dust mass surface density \citep[$\Sigma\rm_{dust}$;][]{Draine 2014}. The symbols are the same as (a), while the horizontal errorbar represents the standard deviation of $\Sigma\rm_{dust}$. (c) [C {\sc ii}]/CO(3--2) line ratio vs. GALEX FUV flux. The symbols are the same as (b).
\label{fig:fig4}
}
\end{figure}

Combining the [C {\sc ii}] \citep{Kapala 2015} and CO(3--2) \citep{Li 2020} data of the five SLIM field, we derived the pixel-by-pixel radial distribution of the [C {\sc ii}]/CO(3--2) ratio, as shown in the Figure \ref{fig:fig4}(a). For pixels with detection of only one line, we calculated the 3$\sigma$ upper limit of another line and then derived the upper or lower limit of the [C {\sc ii}]/CO(3--2) ratio. We computed the mean ratio including both upper and lower limits, using the software Astronomy SURVival Analysis \citep[ASURV;][]{Lavalley 1992}. According to the KOSMA-$\tau$ PDR model \citep{Storzer 1996}\footnote{https://hera.ph1.uni-koeln.de/$\sim$pdr/}, which solves the chemistry and radiative transfer in PDR region with spherical geometry, [C {\sc ii}]/CO flux ratio decreases for increasing metallicity and gas density \citep{Rollig 2006}. Since the metallicity of M31 decreases with increasing galactocentric radius \citep{Draine 2014}, the high ratio in the center cannot be attributed to metallicity. 
\\

We thus compared the [C {\sc ii}]/CO(3--2) ratio with the dust mass surface density map from \cite{Draine 2014}, as shown in Figure \ref{fig:fig4}(b). To evaluate the significance of this correlation, we calculated the Spearman's rank coefficient $\rho$ using only the pixels with both detections. The resultant $\rho$ = -0.45, with a p-value (the probability under the null hypothesis of random variables) $<$ 0.001. The negative correlation indicates that, in general, higher ratio indeed corresponds to lower gas surface density, if assuming a constant gas-to-dust ratio. We note that the gas-to-dust ratio may be lower in the central region \citep{Li 2019}, which will even strengthen our conclusion. As a result, the high [C {\sc ii}]/CO ratio in the center indicates low gas density in this region, consistent with previous studies \citep{Li 2009, Dong 2016, Li 2019}.
\\

We then analyzed the correlation between FUV flux and [C {\sc ii}]/CO(3--2) ratio. The FUV emission from the M31 center is most likely to arise from old stellar populations \citep{Brown 1998, Dong 2015, Viaene 2017}, while the SED fitting method adopted by \cite{Kapala 2017} to derive the attenuated FUV flux is based on the young Galactic disk stellar population. Hence this method is not suitable for the central region. We thus simply used the GALEX FUV flux of each field, and found it positively correlated with the ratio, as indicated in Figure \ref{fig:fig4}(c), with Spearman's rank coefficient $\rho$ = 0.39 (p $<$ 0.001). This correlation suggests that the strong FUV radiation in the center is at least partly responsible for the high ratio. 
\\

We examined the [C {\sc ii}]/[O {\sc iii}] flux ratio, and found it ranges from 0.2 to 4.6 along the nuclear spiral, with an average value of $\sim$1. The [C {\sc ii}]/[O {\sc i}] flux ratio ranges from 0.5 to 11.6 (with only $\sim$7\% pixels $<$1) and has a mean value of 2.1 in the nuclear spiral region, similar to that found in the disk fields \citep[$\sim$2.2,][]{Kapala 2015}, which indicates [C {\sc ii}] is indeed the dominant coolant of neutral gas here. 
As a result, the [C {\sc ii}]/FIR ratio can be considered as an approximation of the photoelectric heating efficiency. Previous studies found this ratio in the range of 0.1$\% \sim 5\%$ \citep{Bakes 1994, Weingartner 2001}, and varies dramatically in different environments. 
\cite{Kapala 2015} found that the [C {\sc ii}]/TIR ratio increases with galactocentric radius using the measurements of the five SLIM fields in the disk of M31. 
\\

\begin{figure}
\centering
\includegraphics[width=3.5in]{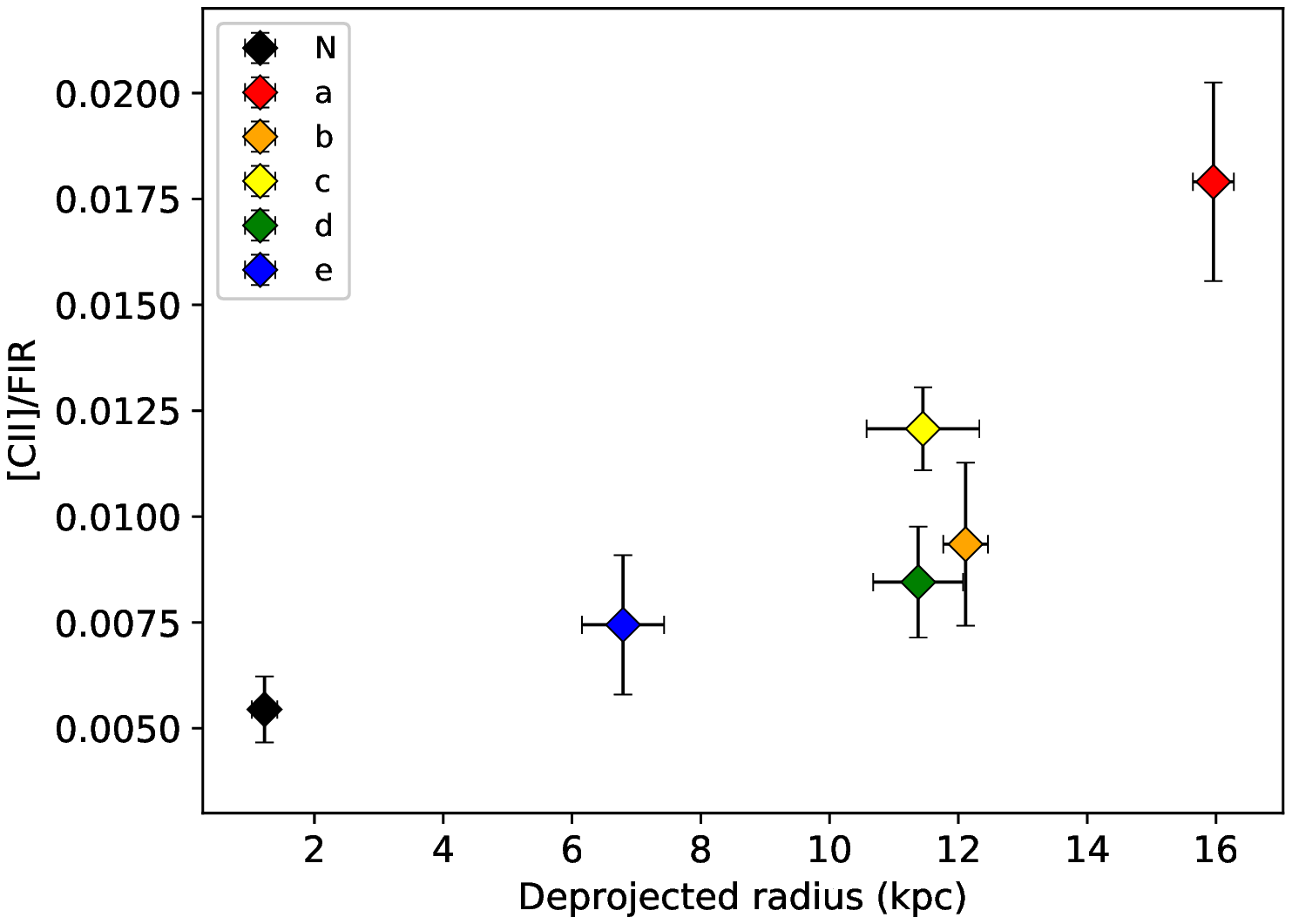}
\includegraphics[width=3.5in]{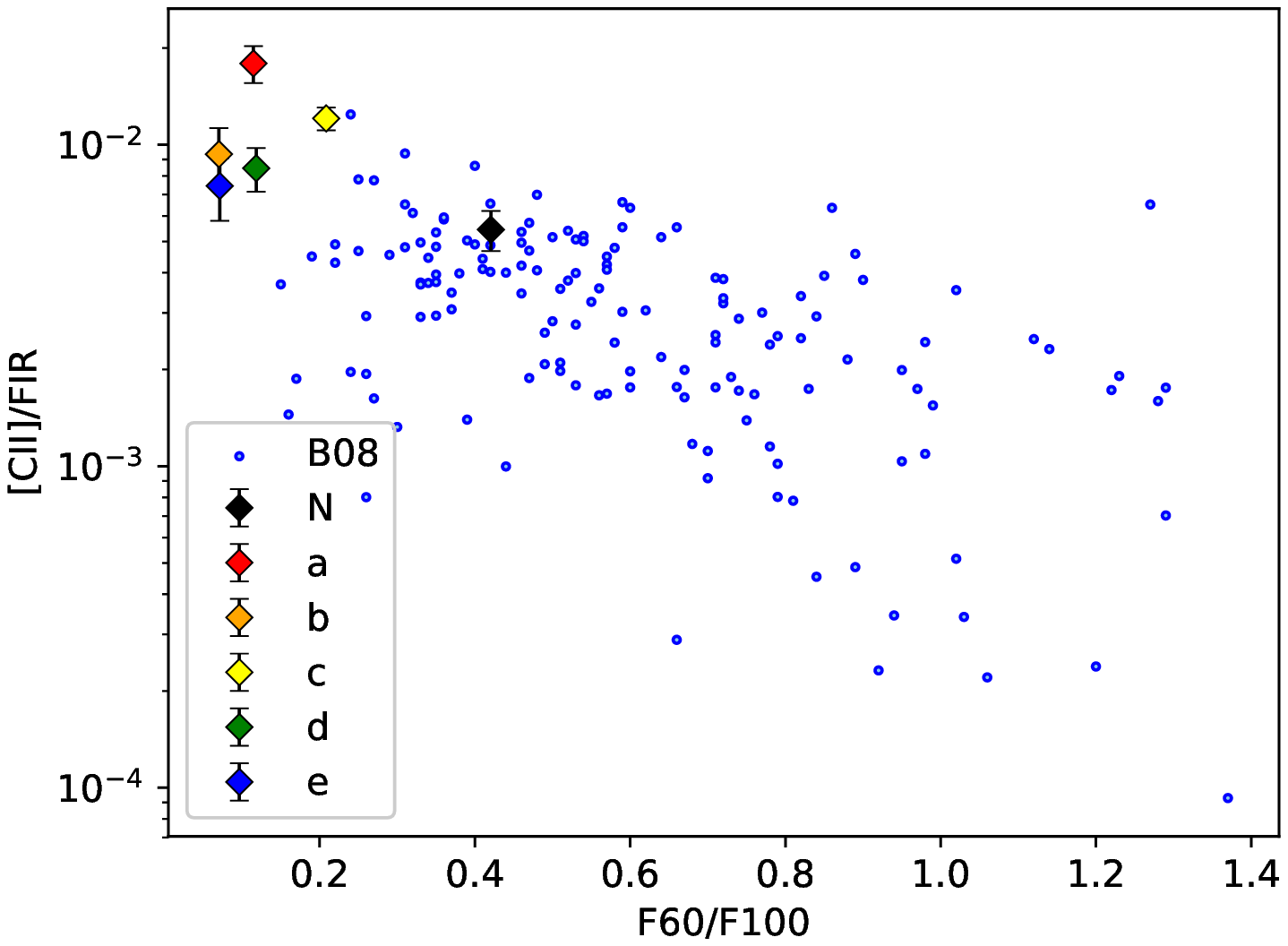}
\caption{$Left$: The radial distribution of mean [C {\sc ii}] to FIR flux ratio (diamonds) of the SLIM fields (denoted by a-e) and the Nucleus field (`N'). The vertical and horizontal error bars represent the uncertainty of the mean ratio and the radial coverage of each field. The FIR map is from \cite{Ford 2013}. 
$Right$: The [C {\sc ii}]/FIR flux ratio plotted against 60 $\micron$/100 $\micron$ continuum flux ratio. The blue points are galaxies from the \cite{Brauher 2008} sample (B08), and the diamonds are the fields of M31. 
\label{fig:fig5}}
\end{figure}

We measured the [C {\sc ii}]/FIR ratio of the central region and the five fields in the disk \citep{Kapala 2015}. The FIR map is from \cite{Ford 2013}, and the [C {\sc ii}] map was convolved to a resolution of 36$\arcsec$ to match the resolution of the FIR map. The ratios are calculated using the mean intensities in each field, as shown in Figure \ref{fig:fig5}. The average [C {\sc ii}]/FIR ratio is 5.4 ($\pm$ 0.8) $\times$ 10$^{-3}$ in the central region, in good agreement with the [C {\sc ii}]/100$\micron$ ratio derived in \cite{Mochizuki 2000}, which is 6 $\times$ 10$^{-3}$. We found that the increasing trend still holds after adding the Nucleus field, as shown in Figure \ref{fig:fig5} left panel. \cite{Kapala 2017} claimed that the increasing trend of [C {\sc ii}]/TIR ratio with radius may be due to the contribution from old stellar population to dust heating increasing towards the center. Consequently, the infrared emission from dust increases, while the gas is not heated efficiently by the less energetic photons from old stellar populations, consistent with previous findings \citep[e.g.][]{Dong 2015, Viaene 2017}.
\\

Previous studies have found a ``deficit'' of the [C {\sc ii}]/FIR ratio at high FIR luminosity \citep[e.g.][]{Malhotra 1997, Stacey 2010} and high 60 $\micron$/100 $\micron$ continuum flux ratio \citep[F60/F100, a proxy for dust temperature; e.g.][]{Malhotra 2001}. Possible reasons for this deficit include inefficient photoelectron ejection of dust grains, due to an increase in grain charge \citep{Malhotra 2001, Croxall 2012} and/or low polycyclic aromatic hydrocarbons fraction \citep{Helou 2001}, and
[C {\sc ii}] saturation at high luminosity \citep[while the FIR increases linearly with FUV radiation;][]{Gullberg 2015}. To investigate this deficit in M31, we calculated the mean F60/F100 ratio of each field using the Infrared Astronomical Satellite (IRAS) 60 and 100 $\micron$ flux. For comparison, we made use of the \cite{Brauher 2008} sample, one of the most complete extragalactic FIR emission line samples, observed by the $ISO$, which contains all morphologies of galaxies. We only used the unresolved galaxy sample. For them, we obtained the FIR flux using the IRAS 60 and 100 $\micron$ flux given in \cite{Brauher 2008} and the formula from \cite{Helou 1988}. The plot of [C {\sc ii}]/FIR ratio and F60/F100 are shown in Figure \ref{fig:fig5} right panel, in which a declining trend is evident. The five SLIM fields of M31 disk fall in the high [C {\sc ii}]/FIR, low F60/F100 regime, showing no obvious deficit, consistent with the overall quiescent and cold environment. The Nucleus field, on the other hand, with a significantly higher dust temperature than the disk \citep{Draine 2014}, exhibits a lower [C {\sc ii}]/FIR ratio. 
\\

\section{Discussion}

The fraction of [C {\sc ii}] coming from ionized gas can affect our discussions above. This fraction is less than 50\% in the five SLIM fields, according to the estimation of \cite{Kapala 2015}, with a typical value of $\sim$20\%. We note that in the central region, due to higher FUV intensity and dust temperature \citep{Draine 2014}, the fraction could be higher \citep{Croxall 2012}. However, the [C {\sc ii}] emission from different components is hard to disentangle due to the missing of [N {\sc ii}] observations and low velocity resolution of the spectra. If we adopt the fraction of [C {\sc ii}] coming from ionized gas to be 53\% \citep[typical for normal spirals;][]{Lapham 2017} for the central region and 20\% for the SLIM fields, the [C {\sc ii}]/CO(3--2) ratio is still higher in the center. As for the [C {\sc ii}]/FIR ratio, subtracting the fraction of [C {\sc ii}] from ionized gas will lower this ratio in the center and enhance the increasing trend with radius.
\\

In addition to the dependency on gas density and FUV intensity, the abundance of different phases of carbon can also be affected by cosmic rays (CRs). CO embedded deep in molecular clouds that are shielded against UV photons can still be effectively destroyed by CRs via He$^+$, leaving the carbon in either C or C$^{+}$ phase \citep{Bisbas 2015}. As suggested by \cite{Bisbas 2019}, high level of CRs can result in decreasing of CO abundance, while [C {\sc ii}] abundance is barely affected or slightly increased. 
We note that in high density ($\sim10^5$ cm$^{-3}$) and high kinetic temperature ($\gtrsim$100 K) environments in the Milky Way, CO brightness can increase with CR ionization rate \citep{Pellegrini 2009}. This could be due to efficient destruction of the He$^+$ that is  otherwise responsible for CO dissociation in these environments, while in average ISM conditions as well as in the M31 center, where the density and temperature are relatively low, CO can still be destroyed via He$^+$ \citep{Bisbas 2017}. 
\\

\begin{figure}
\centering
\includegraphics[width=3.5in]{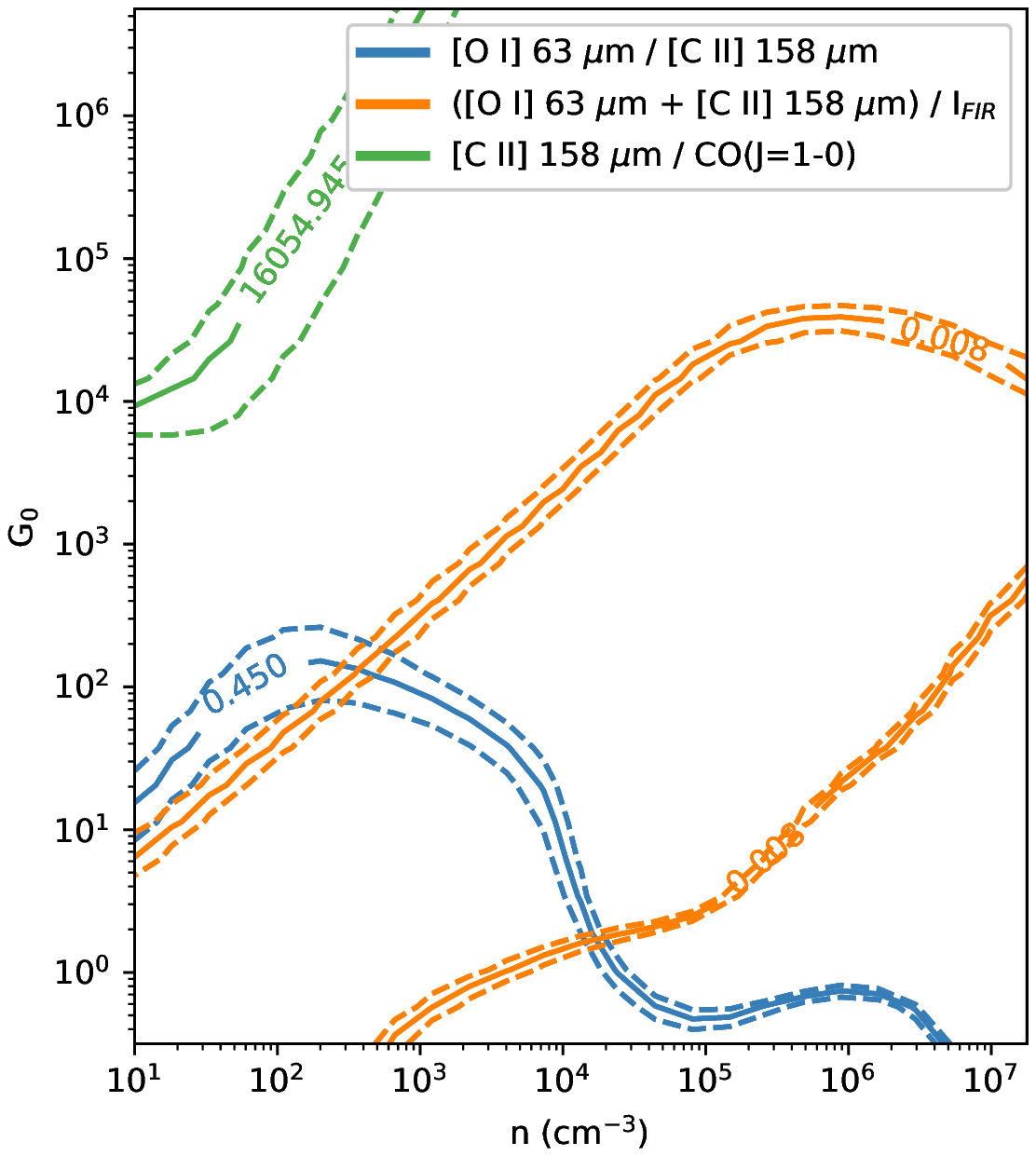}
\includegraphics[width=3.5in]{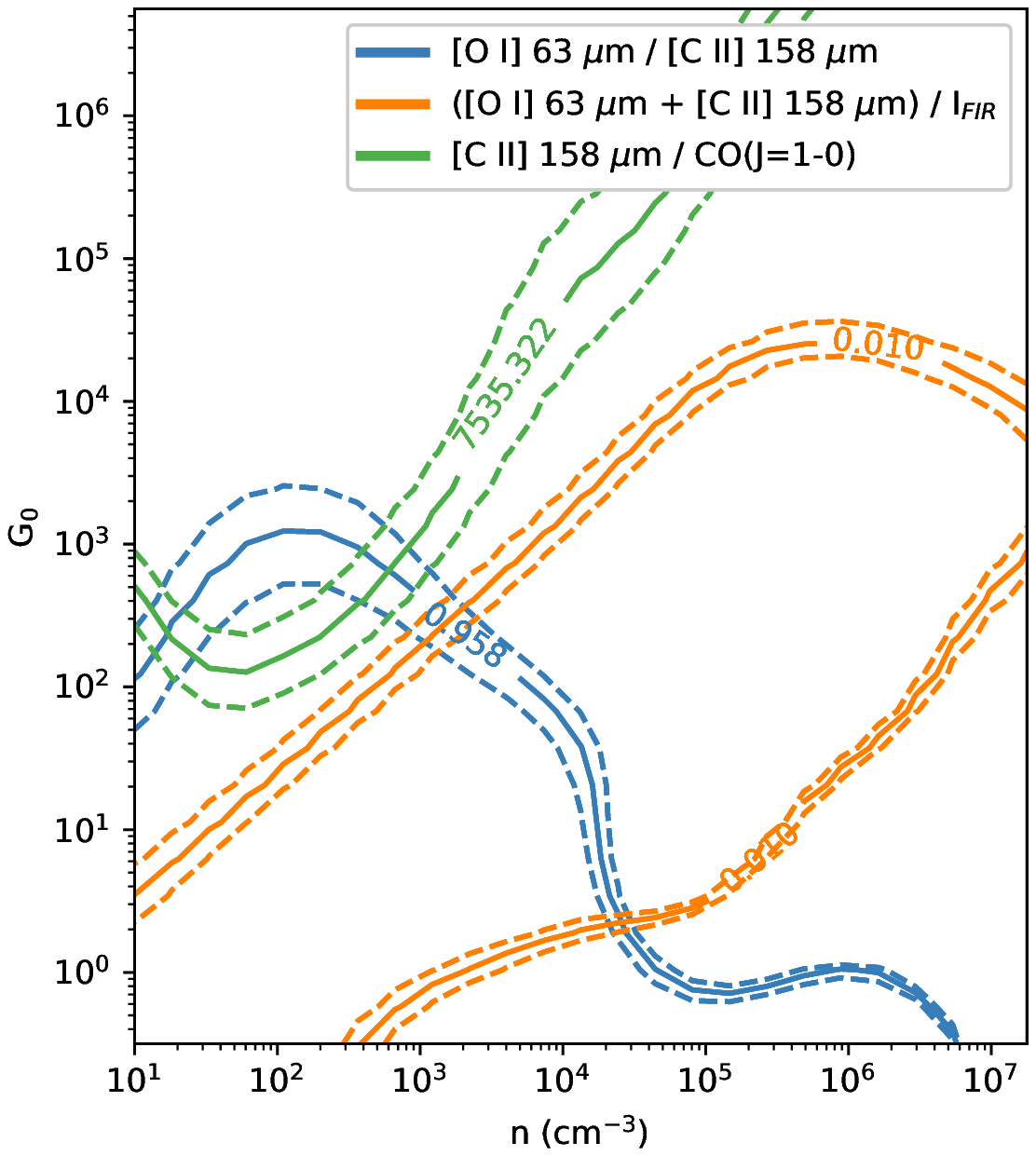}
\caption{Line ratios on the $G_0$ - $n$ plane created by PDRT. Solid lines represent the mean ratios of the nuclear spiral, and dashed lines represent 1$\sigma$ error. The left and right panels are calculated using uncorrected and corrected [C {\sc ii}] and FIR flux, respectively.
\label{fig:fig6}}
\end{figure}

We also derived the physical properties in the central region using the PDR model in the Photo Dissociation Region Toolbox \citep[PDRT;][]{Pound 2008, Kaufman 2006}\footnote{http://dustem.astro.umd.edu/pdrt/}. The model assumes a 2D slab of ISM, and solves the chemistry, radiative transfer and thermal balance in the PDR. Using the [O {\sc i}]/[C {\sc ii}], ([O {\sc i}]+[C {\sc ii}])/FIR and [C {\sc ii}]/CO(1--0) line ratios as diagnostics, the model can constrain gas density $n$ (in units of cm$^{-3}$) and incident FUV intensity $G_0$ \citep[in units of the local interstellar field value, 1.6$\rm \times~10^{-3}~erg~cm^{-2}~s^{-1}$;][]{Habing 1968}, as shown in Figure \ref{fig:fig6}. Here we assume a CO(3--2)/CO(1--0) line ratio of unity, according to \cite{Li 2019}.\\

Since the model only considers the PDR, [C {\sc ii}] from ionized gas should be subtracted, so we corrected the [C {\sc ii}] emission by assuming that 53\% of the [C {\sc ii}] is from ionized gas \citep{Lapham 2017}. In addition, the model assumes the cloud is illuminated by FUV emission from one side, while in the optically thin case, the emission is from both sides, so we divided the FIR flux by a factor of 2, as suggested by \cite{Kaufman 1999}. We also calculated the ratios using the uncorrected [C {\sc ii}] and FIR flux, as shown in the left panel of Figure \ref{fig:fig6}. The intersections of the curves indicate the best-fit values. The minimum $\chi^2$ distribution is at ($n$, $G_0$) = (1.78$\times10^2$, $10^2$) and ($n$, $G_0$) = ($10^3$, 3.16$\times10^2$) for the uncorrected and corrected ratios, respectively. \\

The gas density derived from the PDR model is low compared to the Milky Way center \citep[10$^{3.5}\,\rm cm^{-3}$;][]{Oka 2012}, while consistent with previous studies of M31 center \citep[e.g.][]{Dong 2015}. The relatively low gas density is also consistent with the high [C {\sc ii}]/CO(3--2) ratio in this region, as discussed above. 
To have a consistency check of the radiation field intensity of M31 center, we calculated the mean stellar density using the total stellar mass of 1.4$\times10^{10}~\rm M_\odot$ within the central 680 pc \citep{Dong 2015}, resulting in $\sim$ 45 M$_\odot\,\rm pc^{-3}$. This value is approximately two orders of magnitude higher than the local value \citep[$\sim$ 0.1 M$_\odot\,\rm pc^{-3}$;][]{Holmberg 2000}, hence the radiation field intensity $G_0\sim300$  is reasonable. 
The radiation field intensity of M31 center is comparable to other nearby nuclei, such as M33 \citep{Kramer 2020}, NGC 3184 and NGC 628 \citep{Abdullah 2017}, while stronger than the Milky Way disk \citep[$G_0\sim$ 1-30;][]{Pineda 2013}. 
\\

\section{Summary}

We have carried out the first mapping of FIR fine structure lines [C {\sc ii}], [O {\sc i}] and [O {\sc iii}] on the nuclear spiral of M31 with $Herschel$ PACS spectroscopic observations. Our main results are as follows:

\begin{itemize}

\item[(i)] The three lines all exhibit significant emission along the nuclear spiral, while [O {\sc i}] has a lower SNR and appears more patchy than the other two. The morphology of [C {\sc ii}] emission is consistent with CO(3--2) in the southern half of the nuclear spiral, while [O {\sc iii}] is more similar to H$\alpha$ emission arising from ionized gas. 

\item[(ii)] The [C {\sc ii}]/CO(3--2) ratio is higher in M31 center than in the disk. We found that this ratio has a positive correlation with the FUV flux, and a negative correlation with the dust mass surface density.

\item[(iii)] The [C {\sc ii}]/FIR ratio is 5.4 ($\pm$ 0.8) $\times$ 10$^{-3}$ in the central region which is lower than in the disk. This ratio has an increasing trend with galactocentric radius, which could be due to an increasing contribution from old stellar populations to dust heating towards the center.

\end{itemize} 

Future observations of the [C {\sc i}] line would enable us to determine the abundance of carbon in different phases, and also the fraction of the molecular gas that cannot be traced by CO (the so-called CO-dark gas) in the circumnuclear region. Furthermore, combining the FIR line observations with our optical integrated field spectroscopy (Z. Li et al. in preparation), we will be able to build a comprehensive view of the kinematics and ionization mechanisms of the multiphase ISM in this region.


\begin{acknowledgments}

Herschel is an ESA space observatory with science instruments provided by European-led Principal Investigator consortia and with important participation from NASA. The James Clerk Maxwell Telescope is operated by the East Asian Observatory on behalf of The NationalAstronomicalObservatory of Japan; Academia Sinica Institute of Astronomy and Astrophysics; the Korea Astronomy and Space Science Institute; Center for Astronomical Mega-Science (as well as the National Key Research and Development Program of China with No. 2017YFA0402700). Additional funding support is provided by the Science and Technology Facilities Council of the United Kingdom and participating universities in the United Kingdom and Canada. 
ZNL and ZYL acknowledge support by the National Key Research and Development Program of China (2017YFA0402703) and National Natural Science Foundation of China (grant 11873028). YG acknowledges funding from National Key Research and Development Program of China (Grant No. 2017YFA0402704), National Natural Science Foundation of China Grant No. 11861131007 and 11420101002, and Chinese Academy of Sciences Key Research Program of Frontier Sciences (Grant No. QYZDJ-SSW-SLH008). MWLS acknowledges funding from the UK Science and Technology Facilities Council consolidated grant ST/K000926/1.

\end{acknowledgments}



\begin{thebibliography}{}

\bibitem[Abdullah et al.(2017)]{Abdullah 2017}Abdullah, A., Brandl, B. R., Groves, B., et al. 2017, ApJ, 842, 4
\bibitem[Abel(2006)]{Abel 2006}Abel, N. P. 2006, MNRAS, 368, 1949
\bibitem[Bakes \& Tielens(1994)]{Bakes 1994}Bakes, E. L. O., \& Tielens, A. G. G. M. 1994, ApJ, 427, 822
\bibitem[Bisbas et al.(2015)]{Bisbas 2015}Bisbas, T. G., Papadopoulos, P. P., \& Viti, S. 2015, ApJ, 803, 37
\bibitem[Bisbas et al.(2017)]{Bisbas 2017}Bisbas, T. G., van Dishoeck, E. F., Papadopoulos, P. P., et al. 2017, ApJ, 839, 90
\bibitem[Bisbas, Schruba \& van Dishoeck(2019)]{Bisbas 2019}Bisbas, T. G., Schruba, A., \& van Dishoeck, E. F. 2019, MNRAS, 485, 3097
\bibitem[Brauher et al.(2008)]{Brauher 2008}Brauher, J. R., Dale, D. A.,\& Helou, G. 2008, ApJS, 178, 280
\bibitem[Braun et al.(2009)]{Braun 2009}Braun R., Thilker D. A., Walterbos R. A. M., Corbelli E., 2009, ApJ, 695, 937
\bibitem[Brinks (1984)]{Brinks 1984}Brinks E., 1984, . PhD thesis, Univ. Leiden, Leiden
\bibitem[Brown et al.(1998)]{Brown 1998}Brown, T. M., Ferguson, H. C., Stanford, S. A., \& Deharveng, J.-M. 1998, ApJ, 504, 113
\bibitem[Ciardullo et al.(1988)]{Ciardullo 1988}Ciardullo R., Rubin V. C., Ford W. K., Jr., Jacoby G. H., Ford H. C., 1988, AJ, 95, 438
\bibitem[Croxall et al.(2012)]{Croxall 2012}Croxall, K. V., Smith, J. D., Wolfire, M. G., et al. 2012, ApJ, 747, 81
\bibitem[De Looze et al.(2014)]{De Looze 2014}De Looze, I., Cormier, D., Lebouteiller, V., et al. 2014, \aap, 568, A62
\bibitem[Devereux et al.(1994)]{Devereux 1994}Devereux, N. A., Price, R., Wells, L. A., \& Duric, N. 1994, AJ, 108, 1667
\bibitem[Dong et al.(2015)]{Dong 2015}Dong, H., Li, Z., Wang, Q. D., et al. 2015, MNRAS, 451, 4126
\bibitem[Dong et al.(2016)]{Dong 2016}Dong H., Li Z.,Wang Q. D., Lauer T. R., Olsen K. A. G., Saha A., Dalcanton J. J., Groves B. A., 2016, MNRAS, 459, 2262
\bibitem[Draine et al.(1978)]{Draine 1978}Draine, B. T. 1978, ApJS, 36, 595
\bibitem[Draine et al.(2014)]{Draine 2014}Draine B. T., Aniano, G., Krause, O., et al., \ 2014, ApJ, 780, 172
\bibitem[Ford et al.(2013)]{Ford 2013}Ford G. P. et al., 2013, ApJ, 769, 55
\bibitem[Groves et al.(2012)]{Groves 2012}Groves B. et al., 2012, MNRAS, 426, 892
\bibitem[Gullberg et al.(2015)]{Gullberg 2015}Gullberg, B., De Breuck, C., Vieira, J. D., et al. 2015, MNRAS, 449, 2883
\bibitem[Habing(1968)]{Habing 1968}Habing, H. J. 1968, BAN, 19, 421
\bibitem[Heiles(1994)]{Heiles 1994}Heiles, C. 1994, ApJ, 436, 720
\bibitem[Helou et al.(1988)]{Helou 1988}Helou, George, et al., 1988, ApJS, 68:151-172
\bibitem[Helou et al.(2001)]{Helou 2001}Helou, G., Malhotra, S., Hollenbach, D. J., Dale, D. A., \& Contursi, A. 2001, ApJL, 548, L73
\bibitem[Herrera-Camus et al.(2015)]{Herrera-Camus 2015}Herrera-Camus, R., Bolatto, A. D., Wolfire, M. G., et al. 2015, ApJ, 800, 1
\bibitem[Holmberg \& Flynn(2000)]{Holmberg 2000}Holmberg, J., \& Flynn, C. 2000, MNRAS, 313, 209
\bibitem[Jacoby, Ford \& Ciardullo(1985)]{Jacoby 1985}Jacoby G. H., Ford H., Ciardullo R., 1985, ApJ, 290, 136
\bibitem[Kapala et al.(2015)]{Kapala 2015}Kapala M. J. et al., 2015, ApJ, 798, 24
\bibitem[Kapala et al.(2017)]{Kapala 2017}Kapala, M. J., Groves, B., Sandstrom, K., et al. 2017, ApJ, 842, 128
\bibitem[Kaufman et al.(2006)]{Kaufman 2006}Kaufman, M. J., Wolfire, M. G., \& Hollenbach, D. J. 2006, ApJ, 644, 283
\bibitem[Kaufman et al.(1999)]{Kaufman 1999}Kaufman, M. J., Wolfire, M. G., Hollenbach, D. J., \& Luhman, M. L. 1999, ApJ, 527, 795
\bibitem[Kessler et al.(1996)]{Kessler 1996}Kessler M.F., Steinz J.A., Anderegg M.E., et al., 1996, \aap 315, L27
\bibitem[Kramer et al.(2020)]{Kramer 2020}Kramer, C., Nikola, T., Anderl, S., et al., 2020, arXiv:2005.03683
\bibitem[Lapham et al.(2017)]{Lapham 2017}Lapham, R. C., Young, L. M., \& Crocker, A. 2017, ApJ, 840, 51
\bibitem[Lavalley \& Feigelson (1992)]{Lavalley 1992}Lavalley, M., Isobe, T., \& Feigelson, E. 1992, Astronomical Data Analysis Software and Systems I, 245
\bibitem[Li et al.(2011)]{Li 2011}Li Z., Garcia M. R., Forman W. R., Jones C., Kraft R. P., Lal D. V., Murray S. S., Wang Q. D., 2011, ApJ, 728, L10
\bibitem[Li et al.(2020)]{Li 2020}Li Z., Li Z., Smith M.W.L., Wilson C.D., Gao Y., et al. 2020, MNRAS, 492, 195
\bibitem[Li et al.(2019)]{Li 2019}Li Z., Li Z., Zhou P., Gao Y., Jiang X.-J., Dong H., 2019, MNRAS, 484, 964
\bibitem[Li et al.(2009)]{Li 2009}Li, Z., Wang, Q. D., \& Wakker, B. P., \ 2009, MNRAS, 397, 148
\bibitem[Malhotra et al.(1997)]{Malhotra 1997}Malhotra S. et al., 1997, ApJ, 491, L27
\bibitem[Malhotra et al.(2001)]{Malhotra 2001}Malhotra, S., Kaufman, M. J., Hollenbach, D., et al. 2001, ApJ, 561, 766
\bibitem[McConnachie et al.(2005)]{McConnachie 2005}McConnachie A. W., Irwin M. J., Ferguson A. M. N., Ibata R. A., Lewis G. F., Tanvir N., 2005, MNRAS, 356, 979
\bibitem[Melchior \& Combes(2013)]{Melchior 2013}Melchior A. L., Combes F., 2013, \aap, 549, A27
\bibitem[Mochizuki(2000)]{Mochizuki 2000}Mochizuki K., 2000, \aap, 363, 1123
\bibitem[Muraoka et al.(2007)]{Muraoka 2007}Muraoka K. et al., 2007, PASJ, 59, 43
\bibitem[Oka et al.(2012)]{Oka 2012} Oka, T., Onodera, Y., Nagai, M., et al. \ 2012, ApJS, 201, 14
\bibitem[Ott(2010)]{Ott 2010}Ott, S. 2010, in ASP Conf. Ser. 434, Astronomical Data Analysis Software and Systems XIX, ed. Y. Mizumoto et al. (San Francisco, CA: ASP), 139
\bibitem[Pellegrini et al.(2009)]{Pellegrini 2009}Pellegrini, E. W., Baldwin, J. A., Ferland, G. J., Shaw, G., \& Heathcote, S. 2009, ApJ, 693, 285
\bibitem[Petuchowski et al.(1993)]{Petuchowski 1993}Petuchowski, S. J., \& Bennett, C. L. 1993, ApJ, 405, 591
\bibitem[Pilbratt et al.(2010)]{Pilbratt 2010}Pilbratt, G. L., Riedinger, J. R., Passvogel, T., et al. 2010, \aap, 518, L1
\bibitem[Pineda et al.(2013)]{Pineda 2013}Pineda, J. L., Langer, W. D., Velusamy, T., \& Goldsmith, P. F. 2013, A\&A, 554, A103
\bibitem[Pineda et al.(2018)]{Pineda 2018}Pineda J. L., et al., 2018, ApJ, 869, L30
\bibitem[Poglitsch et al.(2010)]{Poglitsch 2010}Poglitsch, A., Waelkens, C., Geis, N., et al. 2010, \aap, 518, L2
\bibitem[Pound \& Wolfire(2008)]{Pound 2008}Pound, M. W., \& Wolfire, M. G. 2008, in ASP Conf. Ser. 394, Astronomical Data Analysis Software and Systems, ed. R. W. Argyle et al. (San Francisco, CA: ASP), 654
\bibitem[Rollig et al.(2006)]{Rollig 2006}Rollig, M., Ossenkopf, V., Jeyakumar, S., Stutzki, J., \& Sternberg, A. 2006, \aap, 451, 917
\bibitem[Sanders \& Mirabel(1996)]{Sanders 1996}Sanders, D. B., \& Mirabel, I. F. 1996, ARA\&A, 34, 749
\bibitem[Stacey et al.(1991)]{Stacey 1991}Stacey, G. J., Geis, N., Genzel, R., et al. 1991, ApJ, 373, 423
\bibitem[Stacey et al.(2010)]{Stacey 2010}Stacey G. J., Hailey-Dunsheath S., Ferkinhoff C., Nikola T., Parshley S. C., Benford D. J., Staguhn J. G., Fiolet N., 2010, ApJ, 724, 957
\bibitem[St$\rm\ddot{o}$rzer et al.(1996)]{Storzer 1996}St$\rm\ddot{o}$rzer H., Stutzki J. \& Sternberg A. 1996, A\&A, 310, 592
\bibitem[Thilker et al.(2005)]{Thilker 2005}Thilker, D. A., Hoopes, C. G., Bianchi, L., et al. 2005, ApJL, 619, L67
\bibitem[Tielens \& Hollenbach(1985)]{Tielens 1985}Tielens, A. G. G. M., \& Hollenbach, D. 1985, ApJ, 291, 722
\bibitem[van den Bergh(1969)]{van den Bergh 1969}van den Bergh S., 1969, ApJS, 19, 145
\bibitem[Viaene et al.(2017)]{Viaene 2017}Viaene S. et al., 2017, \aap, 599, A64
\bibitem[Weingartner \& Draine(2001)]{Weingartner 2001}Weingartner, J. C., \& Draine, B. T. 2001, ApJS, 134, 263
\bibitem[Wilson et al.(2009)]{Wilson 2009}Wilson C. D. et al., 2009, ApJ, 693, 1736

\end{thebibliography}
\end{document}